\documentclass[journal=jacsat,manuscript=article]{achemso}
\usepackage[utf8]{inputenc}
\usepackage{textcomp}
\usepackage{graphicx}
\usepackage[version=4]{mhchem}
\usepackage{siunitx}
\DeclareSIUnit\au{a.u.}
\usepackage{comment}
\usepackage{epstopdf}
\AppendGraphicsExtensions{.tiff}
\usepackage{gensymb}
\usepackage{tabularx}
\usepackage{multirow}
\usepackage{xcolor}
\colorlet{review}{black}

\title{Overview of Topics in Electrocatalysis for Sustainability: Reactions, Electrocatalysts, Degradation, and Mitigation}

\author{Varada Purohit}
\email{varadapurohit2025@gmail.com}
\affiliation{Department of Chemistry, University of Dayton, OH 45469, USA}

\author{Avdhoot Datar}
\email{adatar1@udayton.edu}
\affiliation{Department of Chemistry, University of Dayton, OH 45469, USA}


\begin{document}

\begin{abstract}Electrocatalysis provides an avenue for transitioning the global energy dependence from fossil fuels to renewable energy sources. While electrocatalytic reactions are being used for several decades, recently, there is a growing interest for electrocatalytic reactions that are useful from sustainability perspective. The wide industrial applications of these sustainable electrocatalytic processes is largely limited by the degradation of the electrocatalysts. This review begins with an introduction to such reactions, followed by a detailed discussion of the electrocatalysts. Finally we describe the processes that are responsible for the degradation of electrocatalytic activity.
\end{abstract}

\section{Introduction}
Developing sustainable and clean ways to produce energy and fuel is a major focus of the global scientific community today, as it has the potential to significantly reduce \ce{CO2} emissions and secure our future energy. One of the most promising approaches to achieving this goal is through electrochemical conversion processes \cite{sherrell2024, dey2017molecular}. These processes involve using electricity to drive chemical reactions, which offers a route to convert renewable resources into valuable energy carriers or chemicals \cite{petersen2021electrochemical, shao2016recent, overa2022}. The efficacy of these technologies hinges not only on the intrinsic activity and selectivity of the electrocatalysts but, crucially, on their long-term durability under operational conditions \cite{li2019recent, meier2012degradation, RISCH2023101247}.

Electrocatalysis is a process that accelerates an electrochemical reaction \cite{pletcher1984electrocatalysis}. An electrochemical reaction is a combination of reduction and oxidation reactions. Although both of these reactions involve the transfer of electrons and occur at the interface of electrode and electrolyte, the directions of electron transfer are opposite to each other. In a reduction reaction, electrons are transferred from the electrode (cathode) The catalysts are required, as in any chemical reaction, to accelerate the rate of electron transfer by reducing the activation energy of the redox reaction. 

Examples of electrochemical conversion include the splitting of water to produce hydrogen \cite{you2018innovative}, the reduction of \ce{CO2} to generate valuable hydrocarbons such as methane or ethylene \cite{xiaoding1996mitigation, taheri2013co2, overa2022}, and nitrogen fixation for the production of ammonia\cite{liu2019NH3}. Water electrolysis can convert water (\ce{H2O}) into hydrogen (\ce{H2}) and oxygen (\ce{O2}), a process that can provide a clean and sustainable source of hydrogen when powered by renewable electricity. Similarly, \ce{CO2} reduction reactions can help mitigate greenhouse gas emissions by converting excess \ce{CO2} into useful fuels or chemicals, such as carbon monoxide or formic acid. One of the goal is to develop processes where abundant molecules that are readily available in our environment, such as water, \ce{CO2}, and nitrogen, can be efficiently converted into higher-value products, such as clean fuels or chemical feedstock, with minimal environmental impact. In these electrochemical reactions, electrocatalysts play a vital role by enhancing the efficiency of electron transfer, lowering the activation energy of the reactions, and ensuring faster, more effective transformations.

Despite significant advancements in designing highly active and selective electrocatalysts, their stability remains a formidable challenge \cite{geiger2018stability, zhao2022insight}. In regular applications, the electrocatalytic materials are exposed to harsh electrochemical environments, including varying potentials, fluctuating temperatures, and the presence of corrosive species or impurities. These conditions automatically lead to catalyst degradation, which manifests through a variety of complex mechanisms \cite{RISCH2023101247}. Understanding these degradation pathways is important for overcoming the current limitations toward commercialization of these clean energy technologies. 

\begin{figure}[H]
\includegraphics[width=14 cm]{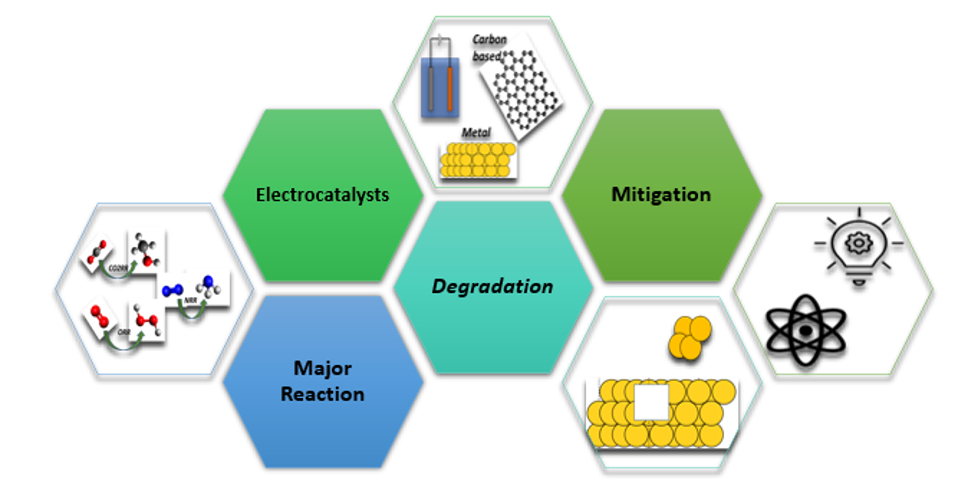}
\caption{Overview of the key topics covered in this review on electrocatalysis for sustainability, including major reactions, catalyst design, degradation mechanisms, and strategies for enhanced performance. \label{overview}}
\end{figure} 

This review aims to provide a concise overview of the field of electrocatalysis in the context of green and sustainable chemistry (Figure \ref{overview}). First, we provide an overview of the key electrochemical reactions that are important from a sustainability perspective. Next, we present a brief survey of commonly used electrocatalysts. Finally, we offer a timely discussion on electrocatalysts' degradation, detailing the underlying mechanisms, the factors influencing these processes, and the emerging strategies aimed at mitigating them.

\section{Types of Electrocatalytic Reactions}
Although a wide range of electrochemical reactions are being investigated for commercial applications, several are particularly vital for advancing sustainability. These reactions play a crucial role in addressing global energy and environmental challenges. In particular, electrocatalytic processes enable the conversion of abundant resources -- such as water, carbon dioxide, and nitrogen -- into valuable fuels and chemicals using renewable electricity. The primary electrocatalytic reactions of interest in this context include:

\subsection{Oxygen Reduction Reaction (ORR)}
The Oxygen Reduction Reaction (ORR) is a key process for various electrochemical devices, most prominently fuel cells (e.g., Proton Exchange Membrane Fuel Cells, PEMFCs) and metal-air batteries (such as Li-air and Zn-air). At the cathode of these devices, ORR converts oxygen into water (or hydroxide ions in alkaline media), enabling the flow of electricity \cite{shao2016recent, dey2017molecular, li2019recent}. Though it is a vital process, the ORR is kinetically sluggish due to its multi-electron transfer nature and the strong oxygen-oxygen bond in the \ce{O2} molecule. The electrocatalysts have to be used to overcome this slow reaction rate and to achieve applicable current densities, which leads to a significant overpotential that limits overall device performance \cite{li2019recent}.

The ORR can proceed via two main pathways: the direct four-electron pathway, yielding water (2\ce{H2O}) in acidic media ($\ce{O2} + 4\ce{H+} + 4\ce{e-} \rightarrow 2\ce{H2O}$) or hydroxide ions (\ce{OH-}) in alkaline media ($\ce{O2} + 2\ce{H2O} + 4\ce{e-} \rightarrow 4\ce{OH-}$), and the two-electron pathway, producing hydrogen peroxide (\ce{H2O2}) in acidic media ($\text{O}_2 + 2\text{H}^+ + 2\text{e}^- \rightarrow \ce{H2O2}$) or hydroperoxide ions ($\text{HO}_2^-$) in alkaline media ($\ce{O2} + \ce{H2O} + 2\ce{e-} \rightarrow \text{HO}_2^- + \text{OH}^-$) \cite{meier2012degradation}. The four-electron pathway is highly desirable for maximizing energy conversion efficiency and avoiding corrosive intermediates like $\text{H}_2\text{O}_2$, which can degrade catalyst materials and membranes \cite{meier2012degradation}. Platinum (Pt) remains the benchmark catalyst for ORR in acidic environments due to its excellent activity and selectivity towards the four-electron pathway; however, its high cost and limited availability are significant barriers to widespread commercialization, driving the search for more sustainable alternatives \cite{shao2016recent, li2019recent}.

\subsection{Hydrogen Evolution Reaction (HER)}
The Hydrogen Evolution Reaction (HER) is a critical cathodic reaction in water electrolysis, which offers a promising pathway for producing clean hydrogen fuel by following electrochemical reactions: in acid media $2\text{H}^+ + 2\text{e}^- \rightarrow \text{H}_2$, and in alkaline media by $2\text{H}_2\text{O} + 2\text{e}^- \rightarrow \text{H}_2 + 2\text{OH}^-$ \cite{zhu2019recent, eftekhari2017electrocatalysts}. As a clean and sustainable energy carrier, hydrogen is envisioned as a key alternative to fossil fuels for the future fuel economy. However, the efficiency of water splitting, and thus the cost-effectiveness of hydrogen production, heavily relies on the performance of HER electrocatalysts.

While platinum (Pt) exhibits the lowest overpotential and fastest kinetics for HER, its high cost and scarcity hinder large-scale implementation \cite{eftekhari2017electrocatalysts}. Consequently, extensive research is dedicated to developing highly efficient, earth-abundant, and durable non-noble metal catalysts, as well as metal-free materials \cite{zhu2019recent}. The HER mechanism typically involves initial proton adsorption, followed by either a Volmer-Heyrovsky or Volmer-Tafel step\cite{bhardwaj2008uncoupled}. The optimization of these steps on the catalyst surface is crucial for high activity. Challenges include balancing activity with long-term stability and reducing the overpotential to make the process economically viable \cite{zhu2019recent}.

\subsection{Hydrogen Oxidation Reaction (HOR)}
The Hydrogen Oxidation Reaction (HOR) is the anodic reaction in fuel cells, particularly in PEMFCs, where hydrogen fuel is oxidized to produce protons and electrons ($\text{H}_2 \rightarrow 2\text{H}^+ + 2\text{e}^-$) \cite{bullock2015molecular, yao2022electrocatalytic}. While the HOR is generally considered to be kinetically fast on platinum in acidic media, making the cathode (ORR) the primary performance bottleneck, its kinetics is found to be significantly slower in alkaline media \cite{yao2022electrocatalytic, cong2018hydrogen}. This presents a major challenge for the development of anion-exchange membrane fuel cells (AEMFCs), which offer the advantage of potentially utilizing non-noble metal catalysts for ORR.

The sluggish kinetics of HOR in alkaline environments necessitates higher loadings of noble metal catalysts or the development of high-performance noble metal-free catalysts for the anode \cite{yao2022electrocatalytic, cong2018hydrogen}. The exact mechanisms of alkaline HOR are still under debate, further complicating the rational design of electrocatalysts for this reaction \cite{yao2022HOR}. Understanding the mechanistic insights, particularly the roles of various adsorbed intermediates and the influence of the electrolyte, is crucial for improving the performance of HOR catalysts. This is essential for unlocking the full potential of AEMFCs and other alkaline fuel cell technologies \cite{yao2022electrocatalytic, shao2011palladium}.

\subsection{Oxygen Evolution Reaction (OER)}
The Oxygen Evolution Reaction (OER) is the anodic half-reaction of water splitting, producing oxygen ($2\text{H}_2\text{O} \rightarrow \text{O}_2 + 4\text{H}^+ + 4\text{e}^-$) \cite{wang2021transition, tahir2017electrocatalytic}. It is also critical for rechargeable metal-air batteries, where it facilitates the regeneration of the metal anode \cite{Lee2016, Zhang2020}. Despite its importance, OER is kinetically sluggish due to the formation of a strong oxygen-oxygen bond and the involvement of multiple proton-coupled electron transfer steps, leading to high overpotentials \cite{eftekhari2017tuning, tahir2017electrocatalytic}. This inherent kinetic barrier represents a major impediment to the overall efficiency of water electrolyzers and metal-air batteries.

Iridium (Ir) and Ruthenium (Ru) oxides are currently the state-of-the-art OER catalysts, especially in acidic conditions, due to their superior activity \cite{tahir2017electrocatalytic}. However, their high cost, scarcity, and limited long-term stability (particularly Ru dissolution in acidic media) restrict their widespread application \cite{tahir2017electrocatalytic, wu2024effective, wan2024earth}. Consequently, there is a global effort to develop cost-effective, earth-abundant alternatives based on transition metals such as Fe, Co, Ni, and Mn, often in the form of oxides, hydroxides, chalcogenides, or phosphides \cite{xu2023current, hu2023electrocatalytic, li2023high, wang2023role}. The challenge lies in designing materials that can achieve high activity while maintaining robust stability under demanding OER operating conditions \cite{eftekhari2017tuning, tahir2017electrocatalytic}.

\subsection{Carbon Dioxide Reduction Reaction (CO2RR)}
The Electrocatalytic Carbon Dioxide Reduction Reaction (CO2RR) is a highly attractive approach for mitigating rising atmospheric \ce{CO2} levels and generating valuable carbon-based chemicals and fuels (e.g., formate, methane, ethanol, ethylene) from renewable electricity \cite{lei2023recent, perry2020developments}. This technology offers a promising route for artificial carbon recycling, potentially integrating carbon capture, utilization, and storage into sustainable energy cycles. However, the high thermodynamic stability of the \ce{CO2} molecule requires a high overpotential for its activation, which remains a critical challenge.

Another major hurdle for CO2RR is achieving high selectivity towards a specific desired product, as \ce{CO2} can be reduced into various $\text{C}_1$ (e.g., CO, HCOOH, $\text{CH}_3\text{OH}$) or $\text{C}_{2+}$ products (e.g., $\text{C}_2\text{H}_4$, $\text{C}_2\text{H}_5\text{OH}$) \cite{jeyachandran2023cutting, saha2022selectivity, Zhang2020CO2RR}. Furthermore, the CO2RR often competes with the energetically favorable HER, which reduces water to hydrogen and can significantly lower the Faradaic efficiency for carbon products \cite{perry2020developments, saha2022selectivity, wu2023electrocatalyst}. Copper (Cu) and Cu-based materials are particularly interesting for CO2RR due to their unique ability to form $\text{C}_{2+}$ products, but their selectivity and stability still need improvement \cite{raciti2018recent}. Research focuses on developing catalysts that can efficiently activate $\text{CO}_2$ while suppressing HER and controlling product distribution \cite{chen2023development, zhang2020co2}.

\subsection{Nitrogen Reduction Reaction (NRR)}
The Electrocatalytic Nitrogen Reduction Reaction (NRR) offers a sustainable and environmentally friendly alternative to the energy-intensive Haber-Bosch process for ammonia (\ce{NH3}) synthesis \cite{tavella2022nitrogen, wan2019heterogeneous}. Ammonia is a vital chemical for fertilizers and is gaining recognition as a potential carbon-free energy carrier and liquid fuel due to its high energy density and hydrogen content \cite{wan2019heterogeneous, majumder2021rational, Ong2024}. The NRR aims to produce $\text{NH}_3$ from $\text{N}_2$ and water under ambient conditions ($\text{N}_2 + 6\text{H}^+ + 6\text{e}^- \rightarrow 2\text{NH}_3$ in acidic media, or $\text{N}_2 + 6\text{H}_2\text{O} + 6\text{e}^- \rightarrow 2\text{NH}_3 + 6\text{OH}^-$ in alkaline media) \cite{shetty2024exploring}.

Despite its enormous potential, NRR faces significant challenges. The dinitrogen molecule (\ce{N2}) is extremely inert due to its strong triple bond, requiring high energy input for activation \cite{wan2019heterogeneous, qing2020recent}. Moreover, the NRR competes intensely with the facile HER, which typically dominates, leading to low ammonia yields and poor Faradaic efficiencies \cite{wan2019heterogeneous, shetty2024exploring, majumder2021rational}. Developing catalysts that can selectively activate \ce{N2} and facilitate the multi-electron transfer steps while suppressing HER is a major research frontier \cite{majumder2021rational}. Mitigation strategies often involve engineering the catalyst's microenvironment to favor \ce{N2} adsorption and activation over proton reduction \cite{wu2023electrocatalyst}.


\section{Electrocatalysts}
Electrocatalysts are essential materials that initiate and accelerate electrochemical reactions. They play a crucial role in determining the efficiency of these processes. In addition, they influence the selectivity and cost-effectiveness of various energy conversion and storage technologies. Their design involves modifying material properties, including composition, structure, and surface characteristics (such as two-dimensional surfaces\cite{Tang2022}), to optimize interactions with reactants and intermediates. Based on their composition, electrocatalysts can be broadly categorized into noble metal-based, non-noble metal, and metal-free carbon-based materials.

\subsection{Noble Metal-Based Electrocatalysts}
Noble metal-based electrocatalysts, particularly those involving platinum (Pt) and other platinum group metals (PGMs) such as palladium (Pd), iridium (Ir), and ruthenium (Ru), have historically been the gold standard due to their exceptional intrinsic activity and stability for various electrochemical reactions.

\subsubsection{Platinum (Pt)}
Platinum is the benchmark catalyst for ORR in acidic media and HER. The superior electrocatalytic activity of platinum stems from its optimized binding energies for reaction intermediates. However, its scarcity and prohibitive cost pose significant barriers to widespread commercialization in applications like fuel cells \cite{shao2016recent, zhao2022advanced, hou2020platinum}. To address these limitations, research on Pt-based catalysts focuses on maximizing Pt utilization and enhancing its durability.

Alloying Pt with transition metals (e.g., Fe, Co, Ni) has emerged as a key strategy to improve ORR activity and stability. The introduction of these secondary metals modifies the electronic structure of Pt (e.g., shifting the d-band center), which can weaken the adsorption of oxygenated species, thereby enhancing kinetics and mitigating poisoning \cite{li2019recent, wu2013platinum, huang2021advanced, mukoyoshi2025nanoalloys}. Examples like ordered $\text{Pt}_3\text{Co}$ intermetallics demonstrate enhanced activity and durability \cite{huang2021advanced}. The resulting lattice strain and ligand effects within these alloys play a crucial role in optimizing the binding energy of ORR intermediates \cite{huang2021advanced}.

Core-shell is an innovative design that minimizes Pt loading by confining it to the outer shell, while a less expensive core material (e.g., Pd, Au, Ni) provides the structural foundation and modulates the electronic properties of the Pt shell \cite{zhao2022advanced, wu2013platinum}. This maximizes the utilization of precious Pt atoms and enhances the catalytic performance and durability through beneficial strain and ligand effects \cite{zhao2022advanced, wu2013platinum}.

Beyond nanoparticles, research explores Pt-based nanowires, nanotubes, and other porous structures. It is found that such structures show a great potential in improving overall catalyst performance in devices like fuel cells \cite{hou2020platinum}. The greater performance of such morphologies is often attributed to the increase of the electrochemically active surface area.

\subsubsection{Other PGMs (Pd, Ir, Ru)}
Palladium (Pd)-based catalysts are investigated for both HOR and ORR, often as alternatives or alloys with Pt, showcasing promising activity and stability \cite{shao2011palladium}. Iridium (Ir) and Ruthenium (Ru) oxides are state-of-the-art catalysts for the OER due to their high activity, especially in acidic environments \cite{tahir2017electrocatalytic}. However, similar to Pt, their high cost and, particularly for Ru, significant dissolution in acidic media, limit their long-term applicability \cite{tahir2017electrocatalytic, wu2024effective, wan2024earth}. Research in this area involves doping and alloying strategies to enhance their stability \cite{wu2024effective, wan2024earth}.

\subsection{Non-Noble Metal Electrocatalysts}
The high cost and limited abundance of noble metals have spurred intensive research into non-noble metal electrocatalysts, which utilize earth-abundant transition metals (\ce{Fe}, \ce{Co}, \ce{Ni}, \ce{Mn}, \ce{Mo}, \ce{W}) and their compounds. These materials aim to provide cost-effective alternatives with comparable or superior performance, particularly in alkaline media.

\subsubsection{Metal-Nitrogen-Carbon (M-N-C) Catalysts}
These are a highly promising class, especially for ORR, often featuring atomically dispersed metal centers (e.g., Fe, Co) coordinated by nitrogen atoms within a porous carbon matrix. The synergistic interaction between the metal, nitrogen, and carbon support creates highly active sites \cite{kumar2023review}. Fe-N-C catalysts, for instance, have shown impressive initial ORR activity, sometimes rivaling that of Pt in alkaline environments \cite{kumar2023review}. Challenges include achieving high active site density, ensuring long-term stability in harsh conditions (especially acidic), and controlling the exact nature of the active sites \cite{wu2024activity}. They are also explored for NRR and CO2RR \cite{manjunatha2020review}.

\subsubsection{Metal Oxides and Hydroxides}
Transition metal oxides (TMOs) and oxyhydroxides are extensively studied for OER and HER due to their tunable electronic properties and diverse structural possibilities. Examples include spinels (\ce{MFe2O4}, where M=Ni, Co, Fe, Mn) \cite{xu2023current, li2023high, wang2023role}, perovskites \cite{alom2022perovskite}, and other mixed metal oxides \cite{ding2021structural}. Molecular metal oxides (polyoxometalates) show a great potential as electrocatalysts in HER \cite{Zeb2023}. These materials often exhibit good activity in alkaline media, and their performance can be enhanced by optimizing their morphology (e.g., porous structures \cite{jin2019recent}), introducing oxygen vacancies, or doping with other elements \cite{xu2023current, li2023high}. Manganese-based oxides, for example, show promise for ORR and water splitting \cite{wang2023role}.

\subsubsection{Metal Chalcogenides (Sulfides, Selenides)}
Transition metal chalcogenides (TMCs), such as $\text{MoS}_2$, $\text{WS}_2$, and their analogues, are gaining significant attention for HER and OER due to their unique electronic structures and abundant active edge sites \cite{su2023defect, tan2023recent, hu2023electrocatalytic, liu2021electrochemical}. Defect engineering (e.g., creating sulfur vacancies) in two-dimensional TMCs has been shown to enhance their electrocatalytic activity \cite{su2023defect, tan2023recent}. Various selenide has also been investigated for selective CO2RR \cite{Nath2022, Deng2023}.

\subsubsection{Metal Phosphides}
Transition metal phosphides (TMPs) like $\text{MoP}$, $\text{CoP}$, and various nickel-molybdenum phosphides are emerging as highly active and stable catalysts for HER and OER \cite{yang2020recent, hu2023electrocatalytic}. Their metallic conductivity and unique electronic properties make them efficient for hydrogen and oxygen evolution. Hybrid structures, such as CoP-embedded nitrogen and phosphorus co-doped mesoporous carbon nanotubes, demonstrate enhanced HER efficiency \cite{singh2023critical}.

\subsubsection{Metal Carbides and Nitrides}
Transition metal carbides (TMCs) and nitrides exhibit good metallic conductivity and often possess Pt-like electronic structures, making them attractive for various electrocatalytic reactions, including ORR, HER, and OER \cite{wang2021transition}. Examples like $\text{Mo}_2\text{C}$ have shown activity for CO2RR \cite{chen2023development}.

\subsection{Metal-Free Carbon-Based Electrocatalysts}
Metal-free carbon-based electrocatalysts offer a sustainable and cost-effective alternative to metal-containing catalysts, leveraging the earth-abundance, high electrical conductivity (for certain morphologies), and tunable surface properties of carbon materials. Their catalytic activity arises from intrinsic defects within the carbon lattice or from the deliberate incorporation of heteroatoms.

\subsubsection{Heteroatom-Doped Carbons}
Doping carbon nanostructures (e.g., graphene, carbon nanotubes, carbon quantum dots) with non-metal heteroatoms such as nitrogen (N), boron (B), sulfur (S), or phosphorus (P) is a primary strategy to enhance their electrocatalytic activity for ORR, HER, OER, and NRR \cite{cheng2022interfacial, liu2016carbon, zhang2015carbon}. Nitrogen doping is particularly effective, as the different nitrogen configurations (pyridinic-N, graphitic-N, pyrrolic-N) within the carbon lattice alter the charge distribution, creating active sites and improving oxygen adsorption and reduction kinetics \cite{cheng2022interfacial, ma2019review}. Boron and sulfur doping can also introduce active sites and enhance electron transfer \cite{liu2016carbon}.

\subsubsection{Defect-Engineered Carbons}
Beyond heteroatom doping, creating intrinsic defects (e.g., vacancies, topological defects, Stone-Wales defects) within the carbon lattice itself can serve as active sites for various electrocatalytic reactions, including ORR, HER, and OER \cite{jia2023defects}. These defects can modify the local electronic structure and provide sites for reactant adsorption. The challenge lies in precisely controlling the type and density of these defects for optimal performance and stability.

\subsubsection{Hybrid Carbon Materials}
Carbon materials often serve as excellent supports or components in hybrid electrocatalysts, enhancing conductivity, providing large surface areas, and preventing the agglomeration of active metal nanoparticles. Examples include graphene-based electrocatalysts integrated with transition metal compounds for HER and electrocatalytic water splitting \cite{nemiwal2021graphene, Liu20253DGraphene}, or porous carbon architectures in lithium-based batteries \cite{wang2022recent}. Various morphologies of carbon-based anodes are often employed in sodium-ion batteries for efficient performance \cite{Qiu2024}. The interfacial engineering of carbon-based materials with heterogeneous components can create specific interfaces that act as active sites or major reaction sites for various reactions (OER, HER, ORR, CO2RR, NRR) \cite{cheng2022interfacial}.


\section{Mechanisms for Degradation}

Despite significant advancements in their design and synthesis, electrocatalysts are inherently susceptible to degradation, which limits their long-term performance and widespread commercialization. These degradation processes are complex, often interconnected, and can occur due to the harsh electrochemical environment, material instability, or interaction with reaction intermediates and impurities. Understanding these mechanisms is crucial for developing strategies to enhance catalyst durability.

\subsection{Corrosion and Oxidation}

Corrosion and oxidation are ubiquitous degradation pathways that impact both the active catalyst material and its support, particularly under high anodic potentials or during dynamic operational cycles.

\subsubsection{Carbon Support Corrosion}
In fuel cells, especially PEMFCs, the carbon support (e.g., carbon black) for PGM catalysts is highly vulnerable to electrochemical oxidation \cite{wei2022degradation, meier2012degradation, singh2019carbon}. This corrosion occurs predominantly at high potentials (e.g., $>0.8\,V_{\text{RHE}}$) or during transient conditions like start-up and shut-down, where the cathode potential can spike \cite{bodner2018degradation}. The oxidation process converts carbon into carbon oxides (\ce{CO2}, \ce{CO}), leading to a reduction in the electrical conductivity of the catalyst layer, a decrease in the active surface area of the catalyst due to detachment of particles, and ultimately, a decline in fuel cell performance \cite{wei2022degradation, meier2012degradation, singh2019carbon, wei2021recent}. Reactive oxygen species (ROS) such as hydrogen peroxide (\ce{H2O2}) and super-oxide radicals, formed as undesired intermediates during the oxygen reduction reaction (ORR), can also chemically attack the carbon support, accelerating its degradation \cite{meier2012degradation, singh2019carbon}. This loss of support integrity can lead to catalyst migration and agglomeration, further exacerbating performance decay \cite{wei2021recent}. Studies on carbon materials highlight how factors like graphitization degree, surface area, and pore structure influence their corrosion resistance \cite{wei2022degradation, wang2022recent}.

\subsubsection{Metal Oxidation and Dissolution}
Active metal components, particularly noble metals like Pt, Ir, and Ru, undergo oxidation and subsequent dissolution, which are major degradation pathways. For Pt catalysts in fuel cells, the formation of platinum oxides (e.g., $\text{PtO}$, $\text{PtO}_2$) at high potentials (e.g., above $0.8\text{--}0.9\,V_{\text{RHE}}$) followed by their dissolution into the electrolyte as $\text{Pt}^{2+}$ or $\text{Pt}^{4+}$ ions is a primary cause of activity loss \cite{chowdury2024degradation, zhao2022advanced, meier2012degradation, bodner2018degradation, li2015enhanced}. This process reduces the electrochemically active surface area over time. This phenomenon is particularly severe during fuel cell start-up/shut-down cycles due to large potential fluctuations \cite{bodner2018degradation}. Similarly, Ru-based catalysts, highly active for the OER, particularly in acidic media, suffer from significant dissolution \cite{wu2024effective, wan2024earth}. This is often linked to the involvement of lattice oxygen in the OER mechanism, leading to the formation of unstable Ru oxide species that readily dissolve, limiting their long-term stability \cite{wu2024effective, wan2024earth}. Tungsten oxide-based catalysts for water splitting also face degradation due to the formation of less active oxide layers that can detach from the electrode surface \cite{yu2024mechanism}. For $\text{PbO}_2$ electrodes, used in electrocatalytic degradation of organic pollutants, the degradation involves corrosion of the $\text{PbO}_2$ layer itself and the underlying substrate, which is crucial for their long-term effectiveness in wastewater treatment \cite{liu2025comprehensive, singh2023critical}.

\subsection{Leaching of Active Sites}
Leaching refers to the selective dissolution and removal of active metal components from the catalyst structure, leading to a direct reduction in the concentration of available active sites for the desired electrochemical reaction. This mechanism is particularly detrimental for noble metal nanoparticles and atomically dispersed catalysts.

For Pt and PGM nanoparticles, the dissolution of individual metal atoms from the particle surface is a fundamental degradation pathway. This dissolved metal can then re-deposit onto larger, more thermodynamically stable particles in a process known as Ostwald ripening \cite{chowdury2024degradation, zhao2022advanced, meier2012degradation, bodner2018degradation, li2015enhanced}. This leads to an increase in the average particle size and a dramatic decrease in the total electrochemically active surface area over time, significantly reducing the catalyst's performance. Alternatively, particle aggregation can occur, where individual nanoparticles merge, also resulting in a reduction of active surface area and accessibility of reaction sites \cite{chowdury2024degradation, li2015enhanced}. Both Ostwald ripening and aggregation are accelerated at higher temperatures and during potential cycling, which promotes surface mobility and dissolution-reprecipitation events \cite{chowdury2024degradation, bodner2018degradation}.

For Metal-Nitrogen-Carbon (M-N-C) catalysts, which are promising non-precious metal alternatives for the ORR, the demetalation of the active metal sites (e.g., iron from Fe-N-C) is a major cause of activity loss, especially in acidic environments \cite{kumar2023review, wu2024activity}. These catalysts rely on atomically dispersed or nanostructured metal centers (often coordinated by nitrogen) embedded within a carbon matrix. Under acidic and oxidative conditions, these metal centers can leach out into the electrolyte, directly deactivating the catalyst \cite{kumar2023review}. This issue is a significant barrier to their widespread adoption in acidic PEMFCs, requiring substantial efforts in material design to improve the anchoring and stability of these active sites \cite{wu2024activity}. Similarly, the significant dissolution of Ru-based catalysts for OER, particularly in acidic media, exemplifies active site leaching, where the precious metal ions are lost from the catalyst, severely limiting its long-term stability and practical applicability \cite{wu2024effective}.

\subsection{Surface Reconstruction}
Surface reconstruction refers to the dynamic and often irreversible changes in the atomic and electronic structure of the catalyst surface that occur under electrochemical operation. These transformations can significantly alter the catalyst's activity and selectivity, sometimes leading to improved performance, but often resulting in detrimental degradation.

In some instances, surface reconstruction can be beneficial, leading to the in situ formation of the `real' active catalyst during operation. For example, some metal oxides or non-oxide compounds may undergo surface oxidation or amorphization under OER conditions, transforming into highly active amorphous metal (oxy)hydroxide layers that are more catalytically efficient than the pristine material \cite{ding2021structural, jones2024toward}. This highlights that the active species for catalysis is not always the as-synthesized material, but rather a dynamically formed phase. Understanding and controlling these beneficial reconstructions is a key strategy for designing self-optimizing catalysts.

However, surface reconstruction often leads to detrimental changes, resulting in the formation of less active, unstable, or passivating layers that hinder electron transfer and block access to active sites. For example, certain transition metal oxides might develop dense, passivating oxide films under anodic potentials, thereby reducing their catalytic efficiency \cite{ding2021structural}. These unfavorable transformations can lead to a gradual decay in performance over time. The dynamic nature of these changes necessitates sophisticated characterization techniques that can probe the catalyst surface \textit{in situ} and \textit{operando}. Techniques like electrochemical liquid-phase transmission electron microscopy (ec-LPTEM) are crucial for visualizing and understanding these real-time transformations at the nanoscale, providing vital insights for rational catalyst design \cite{shen2023insights, Hu2024, Fratarcangeli2025}. Applications of such in situ techniques for monitoring NRR operations is reviewed in ref \cite{Wu2024characterization}. Without such understanding, designing truly stable catalysts is challenging, as the active species might be transient or structurally unstable.

\subsection{Poisoning by Reaction Intermediates and Impurities}
Catalyst poisoning is a significant degradation mechanism wherein foreign chemical species, either present as impurities in the reactant streams or generated as unwanted byproducts during the electrochemical reaction, irreversibly adsorb onto the active sites of the catalyst. This adsorption blocks the active sites, making them unavailable for the desired reaction and consequently diminishing the catalyst's activity and selectivity.

In fuel cells, the primary catalysts are highly susceptible to poisoning from impurities in the fuel and air feeds. For instance, the Pt-anode catalyst is severely affected by carbon monoxide poisoning \cite{perry2006systems, patil2023degradation}. Even trace amounts of CO strongly adsorb on Pt sites, blocking them and leading to significant performance loss, especially at lower operating temperatures where CO desorption is slow \cite{patil2023degradation, mohtadi2005effect}. Other sulfur-containing compounds (e.g., $\text{H}_2\text{S}$) and hydrocarbons present in impure hydrogen fuel can also adsorb on Pt surfaces, further contributing to poisoning \cite{patil2023degradation, mohtadi2005effect}. On the cathode side, impurities in the air feed, such as \ce{SO2} or \ce{NH3}, can adsorb on the active sites of ORR catalysts or interact with the proton exchange membrane, negatively impacting cell performance and stability \cite{chowdury2024degradation, patil2023degradation, uribe2002effect}.

Beyond external impurities, poisoning can also arise from undesired reaction intermediates. In the ORR, incomplete reduction of oxygen can lead to the formation of ROS like hydrogen peroxide ($\text{H}_2\text{O}_2$) and super-oxide radicals. These species are not only detrimental because they represent inefficient energy conversion (two-electron pathway instead of four-electron) but also because they can chemically attack and degrade both the carbon support and the active metal sites, contributing to catalyst instability \cite{meier2012degradation}.

A critical challenge for multi-electron reactions like the CO2RR and the NRR is the intense competition from the kinetically facile HER\cite{saha2022selectivity, wan2019heterogeneous, majumder2021rational}. In aqueous electrolytes, protons are readily available, and their reduction to hydrogen can preferentially occur on the catalyst surface. The strong adsorption of hydrogen intermediates ($\text{H}_{\text{ad}}$) on the active sites effectively `poisons' them, preventing the adsorption and activation of the less reactive $\text{CO}_2$ or $\text{N}_2$ molecules \cite{saha2022selectivity, majumder2021rational, wu2023electrocatalyst}. This competitive adsorption significantly reduces the selectivity and Faradaic efficiency for the desired carbon or nitrogen products, making it a major hurdle for the practical implementation of CO2RR and NRR \cite{saha2022selectivity, wu2023electrocatalyst}. Mitigation strategies often involve engineering the catalyst's microenvironment to suppress HER or designing sites that selectively bind to $\text{CO}_2$ or $\text{N}_2$ over hydrogen intermediates \cite{wu2023electrocatalyst}.


\section{ Factors Influencing Degradation}
The degradation of electrocatalysts is not solely an intrinsic material property but is significantly influenced by a complex interplay of various external operational conditions and the catalyst's inherent characteristics. Understanding these influencing factors is paramount for designing more robust and durable electrocatalytic systems.

Potential cycling is one of the most critical and well-studied factors influencing degradation, particularly in fuel cells, often referred to as start-stop or load cycling \cite{chowdury2024degradation, meier2012degradation}. During these dynamic operations, the catalyst is exposed to a wide range of potentials. For Pt catalysts in PEMFCs, repeated cycling between high potentials (e.g., $1.0\text{--}1.2\,V_{\text{RHE}}$, characteristic of fuel starvation or air purging during shut-down) and low potentials (e.g., $0.6\,V_{\text{RHE}}$, typical operating potential) severely accelerates degradation \cite{meier2012degradation, bodner2018degradation, li2015enhanced, patil2023degradation}. At high potentials, Pt oxidizes, forming surface oxides; upon potential reversal, these oxides are reduced, but some Pt atoms can dissolve into the electrolyte. This repeated oxidation-reduction-dissolution cycle leads to significant Pt loss, particle agglomeration, and a drastic reduction in the electrochemically active surface area \cite{chowdury2024degradation, meier2012degradation, bodner2018degradation, li2015enhanced}. Beyond Pt, carbon supports are also prone to oxidative corrosion at high potentials, leading to structural collapse and detachment of catalyst particles \cite{wei2022degradation, bodner2018degradation}. The rate and extent of degradation are highly dependent on the upper and lower potential limits, sweep rate, and duration of cycling.

Temperature plays a dual role in influencing catalyst degradation. Generally, elevated operating temperatures accelerate the kinetics of most chemical and electrochemical processes, including detrimental ones like dissolution, corrosion, and Ostwald ripening of catalyst nanoparticles \cite{chowdury2024degradation, patil2023degradation}. Higher temperatures can increase the mobility of surface atoms, facilitating particle growth and agglomeration, which reduces the active surface area \cite{li2015enhanced}. However, very low operating temperatures can also induce degradation, particularly in PEMFCs, where water freezing within the porous electrode structures can cause mechanical stress, cracking, and damage to the catalyst layer and membrane, leading to performance loss \cite{chowdury2024degradation, patil2023degradation}. Therefore, maintaining an optimal temperature window is crucial for catalyst durability.

The pH and composition of the electrolyte significantly impact catalyst stability and the specific degradation pathways. The stability of many transition metal oxides and non-noble metal catalysts is highly sensitive to pH, with some being stable in alkaline environments but prone to dissolution in acidic conditions, and vice versa \cite{alom2022perovskite, wu2024effective}. For instance, the kinetics of the HOR are markedly slower in alkaline media compared to acidic media, influencing the choice and stability of catalysts for anion exchange membrane fuel cells \cite{yao2022electrocatalytic, cong2018hydrogen}. The presence of specific ions in the electrolyte can also affect dissolution rates (e.g., halide ions can promote Pt dissolution) or lead to poisoning \cite{patil2023degradation}. Furthermore, the accumulation of corrosive species or reaction intermediates can lead to localized pH changes within the catalyst layer, creating microenvironment that accelerate degradation even if the bulk electrolyte pH is controlled \cite{patil2023degradation}.

Finally, mass transport limitations within the electrode structure can indirectly contribute to degradation. Inadequate supply of reactants or inefficient removal of products can lead to local concentration gradients. For example, in ORR, poor oxygen transport can result in the accumulation of reactive oxygen species ($\text{H}_2\text{O}_2$) at the cathode, which can then attack the carbon support and the catalyst itself \cite{meier2012degradation}. Similarly, the buildup of gaseous products (e.g., $\text{O}_2$ bubbles in OER or $\text{H}_2$ bubbles in HER) can block active sites, hinder mass transfer, and cause physical damage to the catalyst layer over time. These localized effects create conditions that are more conducive to various degradation mechanisms than the bulk operating environment might suggest \cite{patil2023degradation}. The intrinsic material properties, such as electronic structure, surface facets, defect density, and chemical bonding within the catalyst, ultimately dictate its inherent resistance to these degradation pathways \cite{jia2023defects}.


\section{Mitigation Strategies}
The persistent challenge of electrocatalysts' degradation necessitates innovative strategies that integrate fundamental materials science with advanced engineering principles. Effective mitigation approaches often involve tailoring the catalyst's composition, structure, and local environment to enhance its resilience against degradation mechanisms while maintaining high activity.

\subsection{Composition Engineering}
Catalyst design and compositional engineering is a primary avenue for enhancing durability. For platinum-based catalysts, alloying Pt with transition metals (e.g., Fe, Co, Ni) introduces lattice strain and ligand effects that can modify the electronic structure of Pt, optimizing its binding energies for reaction intermediates and making it less susceptible to dissolution and poisoning \cite{shao2016recent, zhao2022advanced, huang2021advanced, mukoyoshi2025nanoalloys}. These alloys, particularly ordered inter-metallic structures, demonstrate enhanced stability against Pt dissolution and improved ORR kinetics \cite{huang2021advanced}. Core-shell structures represent another elegant solution, where a thin, protective Pt shell encapsulates a cheaper, often non-noble metal core. This design not only maximizes the utilization of precious Pt but also protects the core from dissolution and can tune the electronic properties of the surface Pt atoms, leading to both higher activity and improved durability \cite{shao2016recent, zhao2022advanced, wu2013platinum}. For other metal catalysts, such as Ru-based OER catalysts, strategies like alloying with more stable elements are explored to mitigate their significant dissolution in acidic media \cite{wu2024effective, wan2024earth}.

\subsection{Structure Modifications}
Structural and morphological control also plays a crucial role for enhancing the durability of a electrocatalyst. Designing catalysts with controlled porous structures can improve mass transport kinetics, reduce local concentrations of corrosive species, and provide larger surface areas for active sites, all contributing to enhanced stability \cite{jin2019recent}. Furthermore, the choice of support material is critical. Using highly graphitized carbon supports significantly enhances their corrosion resistance against electrochemical oxidation compared to amorphous carbon, providing a more stable and robust platform for active metal nanoparticles and improving overall catalyst durability in fuel cells \cite{wei2022degradation, wang2022recent, li2015enhanced, qiao2021advanced}. This approach directly addresses the carbon corrosion issue, a major degradation pathway.

\subsection{Interfacial Engineering}
Interfacial engineering focuses on optimizing the interactions at the boundaries between different catalyst components (e.g., metal/support, metal/metal oxide interfaces). Strong electronic and chemical interactions at these interfaces can stabilize active sites, prevent their leaching, and promote efficient electron transfer kinetics \cite{ding2021structural, hou2020platinum, cheng2022interfacial}. This can also be leveraged to induce beneficial surface reconstructions while suppressing detrimental ones, thereby guiding the catalyst towards more stable and active configurations during operation \cite{ding2021structural}. For instance, certain interfaces can stabilize atomically dispersed active sites, preventing their agglomeration or dissolution \cite{wu2024activity}.

\subsection{Microenvironment Engineering}
Controlling the microenvironment involves tailoring the local chemical and physical environment around the active sites to favor desired reactions and suppress degradation pathways. For reactions facing strong competition from the HER, such as NRR and CO2RR, engineering a hydrophobic surface or controlling the local pH can effectively suppress HER by limiting proton access to the active sites, thereby enhancing selectivity and Faradaic efficiency for the target product \cite{wan2019heterogeneous, wu2023electrocatalyst}. This strategy creates a more favorable environment for the specific reactant (e.g., $\text{N}_2$ or $\text{CO}_2$) to adsorb and react.

\subsection{Developing New Materials}
Finally, the development of inherently robust materials and protective coatings offers direct approaches to combating degradation. This includes exploring new classes of stable transition metal compounds like specific phosphides and chalcogenides, or highly stable Metal-Organic Framework (MOF)-derived materials that are intrinsically designed to withstand harsh electrochemical environments \cite{hu2023electrocatalytic, li2023high, nemiwal2021metal}. Applying ultra-thin protective layers or functional coatings over catalyst particles can physically shield them from direct contact with corrosive electrolytes or reactive intermediates while still allowing reactants to access the active sites, thus extending their operational lifetime \cite{cherevko2018stability}. Combined, these multifaceted strategies are essential for bridging the gap between laboratory-scale catalyst performance and real-world durability requirements.

\section{Conclusions}
The widespread adoption of electrocatalysis in sustainable energy technologies hinges critically on overcoming the persistent challenge of catalyst degradation. While tremendous progress has been made in enhancing the activity and selectivity of electrocatalysts, their long-term durability under demanding operational conditions remains a formidable bottleneck. This review has illuminated the diverse and interconnected mechanisms that lead to catalyst degradation, including corrosion and oxidation, leaching of active sites, detrimental surface reconstruction, and poisoning by reaction intermediates and impurities. Each mechanism presents unique challenges that are further exacerbated by various operating factors such as potential cycling, elevated temperatures, and specific electrolyte compositions.

Addressing these complex degradation issues necessitates a concerted and systematic approach across the entire research and development pipeline. First and foremost, stability tests should become a norm for all studies that present new electrocatalysts. Moving beyond initial activity metrics to include rigorous, standardized, and long-term durability assessments under relevant operating conditions is crucial. This paradigm shift will facilitate a more realistic evaluation of catalyst viability, allowing for the prioritization of materials that not only exhibit high initial performance but also possess the inherent robustness required for practical applications.

Secondly, given that degradation mechanisms for each electrocatalyst are not only distinct but also highly dependent on various intrinsic and extrinsic factors, there is an urgent need for developing standardized protocols to assess the temporal stability of electrocatalytic performance. Such standardization would enable truly meaningful and reproducible comparisons across different research groups and materials, significantly accelerating the identification, optimization, and scale-up of durable electrocatalysts. These protocols should ideally simulate real-world operating conditions and include accelerated stress tests to quickly gauge long-term performance.

Finally, to avoid speculative or false assignments of degradation mechanisms, it is imperative that hypotheses regarding these mechanisms are rigorously verified via complementary experimental or theoretical methods. The integration of advanced \textit{in situ} and \textit{operando} characterization techniques (e.g., ec-LPTEM, XAS, operando spectroscopy) with sophisticated computational modeling (e.g., DFT, machine learning) can provide unprecedented atomic-level insights into dynamic catalyst transformations and active site evolution under reaction conditions. This multi-pronged approach will foster a deeper fundamental understanding, allowing for the rational design of more robust and long-lasting electrocatalysts that can withstand the rigors of continuous operation and ultimately drive the energy transition towards a sustainable future.


\section*{Authors' Contributions}VP prepared the original draft of the manuscript. Both the authors (VP and AD) contributed towards conceptualization, writing--review and editing. All authors have read and agreed to the published version of the manuscript.

\section*{Funding}This research received no external funding.

\section*{Acknowledgments}The authors gratefully acknowledge MDPI for the waiver of the article processing charge.

\section*{Conflicts} The authors declare no conflicts of interest.

\bibliography{achemso-demo}

\end{document}